\documentclass[]{article}  

\usepackage{amsmath,amsfonts,amssymb}
\usepackage{graphicx}
\usepackage[colorlinks=true, allcolors=blue]{hyperref}

\usepackage[margin=1.0in]{geometry}
\usepackage{bm}
\usepackage{mathrsfs}
\usepackage{braket}
\usepackage{dsfont}
\usepackage{todonotes}
\usepackage{subfig}

\usepackage[usenames,dvipsnames]{pstricks}
\usepackage{epsfig}
\usepackage{pst-grad} 
\usepackage{pst-plot} 
\usepackage[space]{grffile} 
\usepackage{etoolbox} 
\makeatletter 
\patchcmd\Gread@eps{\@inputcheck#1 }{\@inputcheck"#1"\relax}{}{}
\makeatother

\usepackage{authblk}

\newcommand{\wtphi}{%
	\mspace{2mu}%
	\widetilde{\mspace{-2mu}\smash[t]{\phi}}%
}

\title{Single-shot Adaptive Measurement\\
	for Quantum-enhanced Metrology}

\author[a]{Pantita Palittapongarnpim}
\author[b,c]{Peter Wittek}
\author[a,d,e,f]{Barry C. Sanders}
\affil[a]{Institute for Quantum Science and Technology, University of Calgary, Alberta T2N~1N4 Canada}
\affil[b]{ICFO -- The Institute of Photonic Sciences, 08860 Castelldefels (Barcelona), Spain}
\affil[c]{University of Bor\aa s, 501 90 Bor\aa s, Sweden}
\affil[d]{Program in Quantum Information Science, Canadian Institute for Advanced Research, Toronto, Ontario M5G~1Z8, Canada}
\affil[e]{Hefei National Laboratory for Physical Sciences at Microscale\\ 
	University of Science and Technology of China\\ 
	Hefei, Anhui 230026, People\textsc{\char13}s Republic of China}
\affil[f]{Shanghai Branch, CAS Center for Excellence and Synergetic Innovation Center
	in Quantum Information and Quantum Physics\\ 
	University of Science and Technology of China\\ 
	Shanghai 201315, People\textsc{\char13}s Republic of China}

\begin{document} 
	\maketitle
	
	\begin{abstract}
		
        Quantum-enhanced metrology aims to estimate an unknown parameter such that the precision scales better than the shot-noise bound. 
        Single-shot adaptive quantum-enhanced metrology (AQEM) is a promising approach that uses feedback to tweak the quantum process according to previous measurement outcomes.
        Techniques and formalism for the adaptive case are quite different from the usual non-adaptive quantum metrology approach due to the causal relationship between measurements and outcomes.        
        We construct a formal framework for AQEM by modeling the procedure as a decision-making process, and we derive the imprecision and the Cram\'{e}r-Rao lower bound with explicit dependence on the feedback policy.        
        We also explain the reinforcement learning approach for generating quantum control policies, which is adopted due to the optimal policy being non-trivial to devise.
        Applying a learning algorithm based on differential evolution enables us to attain imprecision for adaptive interferometric phase estimation, which turns out to be SQL when non-entangled particles are used in the scheme.
	\end{abstract}
	
	\section{Introduction}\label{sec:intro}
	Quantum-enhanced metrology studies the use of a quantum resource to obtain an estimate $\wtphi$ of an unknown parameter $\phi$ such that the imprecision exceeds what was set by the use of a classical strategy. This lower bound is known as the standard quantum limit (SQL) and asymptotically reaches the scaling $\Delta\wtphi \propto N^{-1/2}$, where $N$ is the number of resources~\cite{GLM11,TA14}. 
	quantum metrology techniques break this limit by utilizing uniquely quantum mechanical properties, such as squeezing and entanglement \cite{BS13,CSSM14,JKF+09}, to reach the Heisenberg limit (HL), given for interferometric phase estimation as approaching $N^{-1}$~\cite{ZPK12}. 
	This quadratic improvement in precision has been proposed for advancing gravitational wave detectors~\cite{Cav81,AAA+16} and atomic clocks~\cite{BIWH96}. 
	
    Adaptive quantum-enhanced metrology (AQEM) is an approach that is able to achieve quantum-enhanced precision by employing entangled states along with an adaptive procedure~\cite{HBB+07,WBB+09,OIO+12,RPH2015}.
    By dividing the particles into a string of single-particle bundles, the measurement scheme can be made practical using available technology in single-particle detection, although fast feedback control is assumed.
    In this work, we consider a scheme in which the $N$-particle state is divided into bundles of $L$ particles.
    A single-shot AQEM scheme is performed by applying an adaptive procedure that does not repeat over the course of the measurement~\cite{Wis95}.
	
    A key to achieving quantum-enhanced precision using AQEM is the feedback policy, which is a set of instructions governing the adaptive procedure~\cite{Wis95}.
    How the policy determines the imprecision can be explained from two perspectives.
    An adaptive measurement scheme is thought to achieve the minimum imprecision by approximating the theoretical optimal measurement~\cite{WBB+09}.
    In this view, an optimal policy steers the AQEM scheme to approximate this measurement within some error bound.
    Another explanation of the working of AQEM is to consider the procedure as a Bayesian measurement, where the prior for $\phi$ is updated according to the outcomes~\cite{LDT14}.
    In this case, a successful policy can be said to maximize the amount of information about $\phi$ as represented by the prior.
    The Bayesian model is useful as it provides a framework to generate successful policies by minimizing the width of the prior after each measurement\cite{BWB01}.
	
    In this work, we present an alternative approach to modeling the AQEM process, namely, the decision-making approach.
    This approach views the adaptive procedure as making a decision after each measurement based on the outcomes and the feedback policy.
    In this framework, the imprecision is the width of the probability distribution $P(\wtphi|\phi)$, which can either be calculated, if possible, or estimated through multiple applications of the adaptive scheme.
    One advantage of using this approach is that it enables us to represent a policy in a way that can be optimized based only on the final imprecision.
	
	%
    Following this formalism, we derive the Cram\'{e}r-Rao lower bound (CRLB)~\cite{Oli2014}, which quantifies the minimum variance achievable by a parameter estimation scheme.
    The CRLB from this formalism explicitly includes the policy, which means the lower bound can only be given when the optimal policy is known.
    As the policy is not trivial to generate, we employ a reinforcement learning algorithm for the task~\cite{HS10}.
    Although the policy found using such algorithm is not guaranteed to be optimal, the policy can be approximated within an acceptable error bound, and conditions can be placed such that quantum-enhanced precision is delivered~\cite{PWZS16}.
	
    With a feedback procedure in place, it is not straightforward to show whether the lower bound approaches SQL when particles are not entangled, as the independent and identical distribution condition is violated.
    To show the lower bound, we apply a reinforcement learning algorithm to a simulation of adaptive interferometric phase estimation.
    The phase estimation problem is chosen because it is a well studied problem arising from applications to gravitational wave detection~\cite{Cav81,AAA+16} and atomic clocks~\cite{BS13}.
    Hence, the lower bounds for both classical and quantum strategies are well-established~\cite{HBZW12,Bra92,DJK15}.
    We compare the imprecision computed from two input states: the sine state~\cite{BWB01}, which is an entangled state, and a train of non-entangled photons~\cite{PS09}.
		
	This paper is structured as follows.
	In Section~\ref{sec:AQEM}, we describe an AQEM procedure and model this as a decision process. The imprecision for this scheme can be formulated from the distribution given by this model.
	In Section~\ref{sec:CRLB}, we derive the CRLB for an AQEM and consider in particular the special case where the bundles are not entangled. This case leads to the simplification of the CRLB.
	In Section~\ref{sec: learning algorithm}, we discuss the use of reinforcement learning algorithm to generate a feasible policy for an AQEM scheme.
	Section~\ref{sec: phase estimation} describes adaptive phase estimation and the way in which the learning algorithm is applied to the problem.
	The results of applying the optimized policies to adaptive phase estimation using entangled and non-entangled state are shown in Section~\ref{sec:results}.
	
	\section{Adaptive quantum-enhanced metrology}\label{sec:AQEM}
	
	In this section, we describe an adaptive measurement scheme that divides $N$ particles into bundles of equal size. We then conceive this scheme as a decision-making process, allowing us to represent the procedure as a decision tree. The formalism leads to an imprecision and a lower bound that are dependent on the policy.
	
	\subsection{Adaptive measurement procedure}\label{subsec:general AQEM procedure}
	\begin{figure}
		\centering
		\subfloat[][]{
			\centering
			\includegraphics[width=0.45\linewidth]{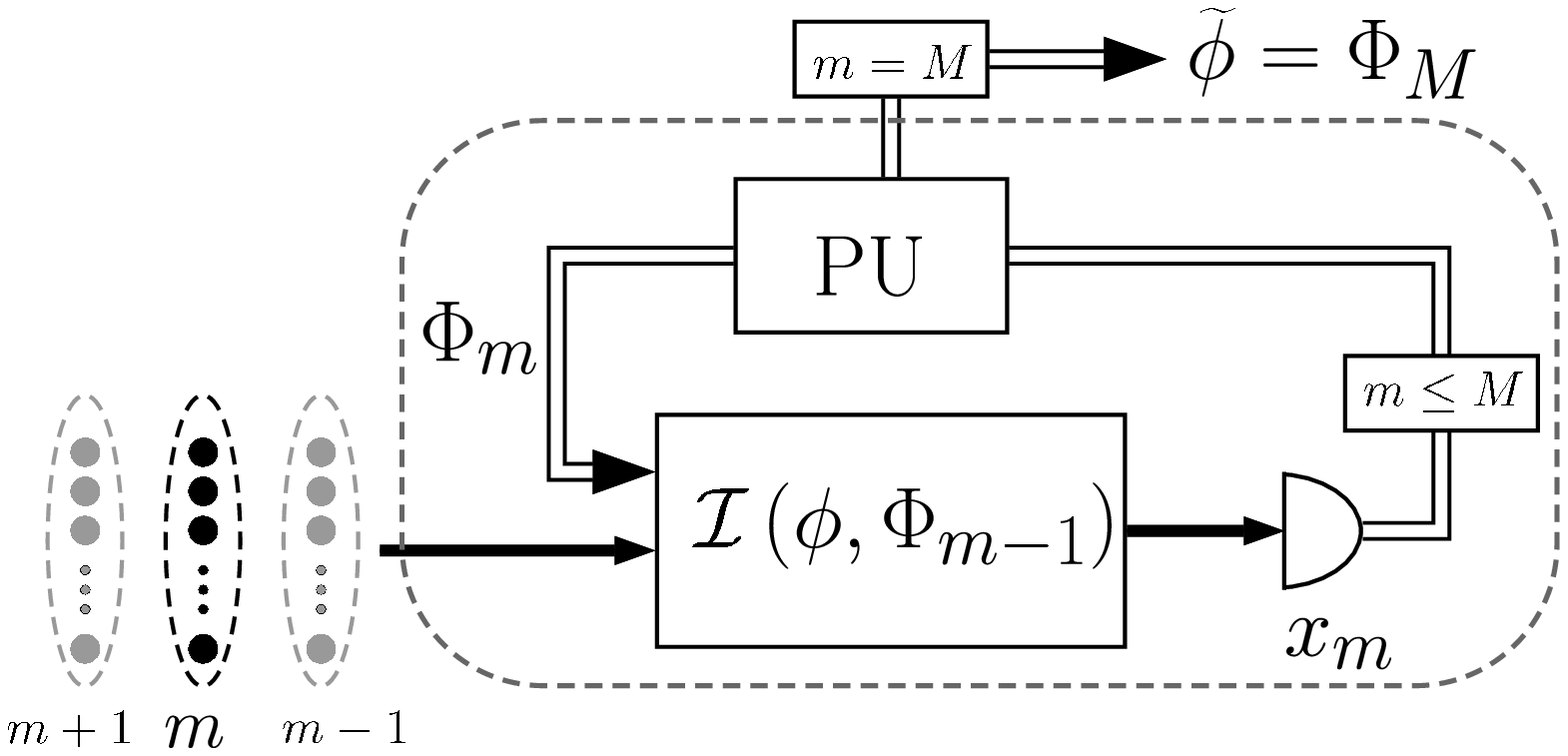}
			\label{fig:aqem}
		}
		\subfloat[][]{
			\centering
						\includegraphics[width=0.45\linewidth]{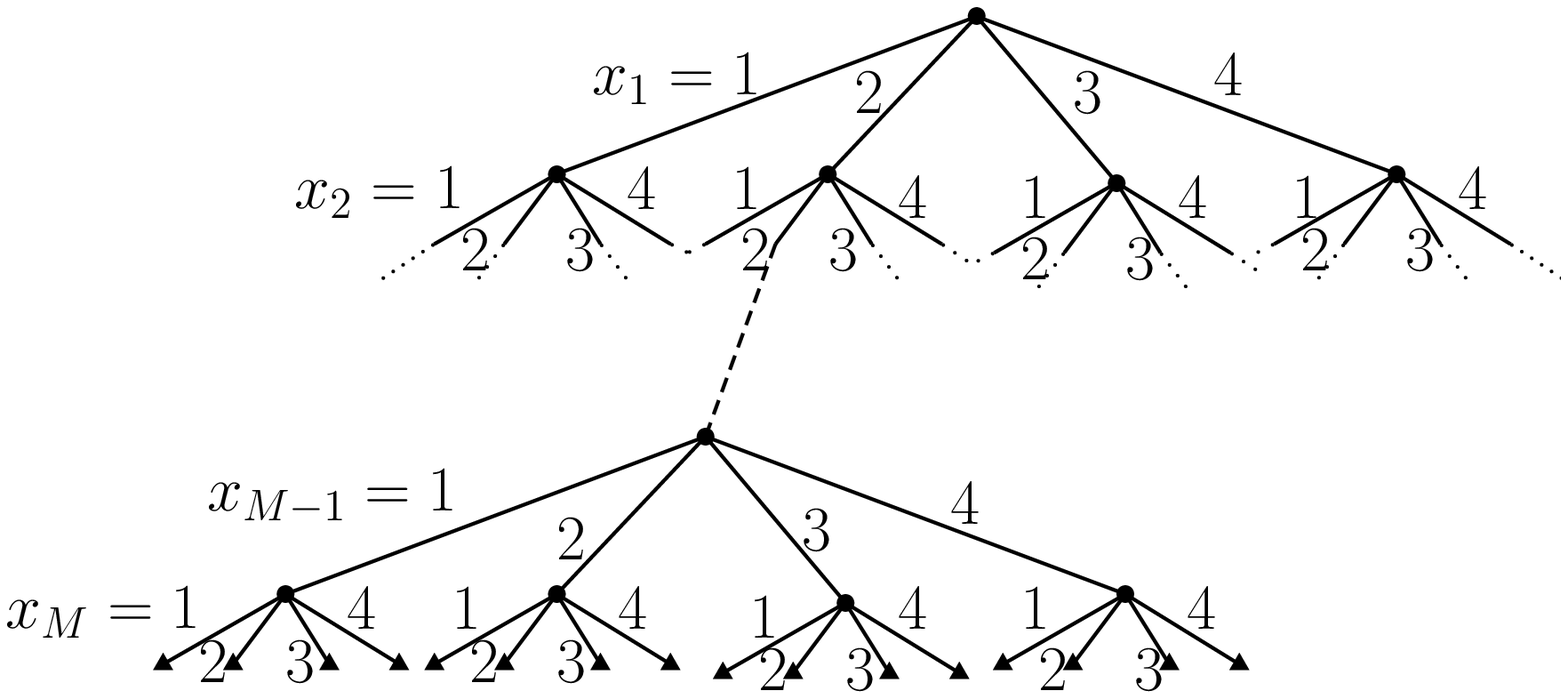}
						\label{fig:decision tree}
		}
			\caption{(a) A diagram of AQEM procedure. 
				The input state $\rho_N$ is divided into~$M$ bundles of $L$ particles,
				each being injected into the process one after another.
				After the detection of the $m^\text{th}$ bundle, the processing unit (PU) uses the history of the measurement outcomes up to that point, $\bm{x}_m=x_1x_2\dots x_m$, in order to determine the control parameter $\Phi_m$, taking the process
				$\mathcal{I}\left(\phi,\Phi_{m-1}\right)\mapsto\mathcal{I}\left(\phi,\Phi_m\right)$ before the $(m+1)^\text{th}$ bundle passes through. 
				This process is repeated until all bundles are measured and $\Phi_M$ is reported as the estimate.
				(b) An example of a decision tree representing an adaptive measurement using $M$ bundles. Each measurement (a node) produces one of the $d^L=4$ possible outcomes represented by a branch.
				The number of leaves grows exponentially with a depth $M$. Here we only expand one of the subtrees at the depth $M-1$ and $M$.}
	\end{figure}	
	In this subsection, we describe a scheme for single-shot quantum-enhanced adaptive measurement (Fig.~\ref{fig:aqem}). The aim of the procedure is to estimate an unknown parameter $\phi$ governing a process $\mathcal{I}$. We consider the case where $N$ $d$-level particles is used as a resource and is divided equally into $M$ bundles of $L$ particles. Each bundle is subjected to the process $\mathcal{I}(\phi,\Phi)$, where $\Phi$ is a controllable parameter. After the $m^{\text{th}}$ bundle is measured, resulting in an outcome $x_m=\{1,\dots,d^L\}$, the value of $\Phi_{m-1}$ is adjusted according to the feedback policy $\varrho$ and the history of outcomes leading up to the $m^\text{th}$ measurement, $\bm{x}_m=x_1x_2\dots x_m$. Here we do not restrict the feedback to be Markovian~\cite{WO12}.    
	The estimate $\wtphi$ is inferred from $\Phi_M(\bm{x}_M)$ once all the bundles are measured, barring any loss of particle.
	
    The $N$ particles are injected into the measurement scheme as bundles of $L$ particles as a strategy to increase the information obtained from the measurement~\cite{Bra94}.
    The bundles can be entangled, and so the state is $\rho_N$, acting on state space $\mathcal{S}\left(\mathscr{H}_d^{\otimes N}\right)$. The particles can be temporally or spatially separated and, hence, can be labeled. In particular, we restrict the input state to those which are symmetric under particle reordering. A permutationally symmetric state is advantageous for an adaptive scheme because the effect of the loss of particle on the imprecision is reduced. In this work, however, we do not include loss.
	
	The $m^\text{th}$ bundle is injected into a process $\mathcal{I}(\phi,\Phi_m)$, which is a quantum channel, i.e., a completely-positive trace-preserving map~\cite{Hol12}. In this formalism, the unitary channel becomes a special case for when no noise or loss is present~\cite{YMK86}. 
	The map is represented by a set of Kraus operators $\{K_i(\phi,\Phi_m)\}$, acting on the subspace $\mathscr{H}_d^{\otimes L}$, such that the condition
	\begin{equation}
		\sum\limits_{i}K_i(\phi,\Phi_m)K_i^\dagger(\phi,\Phi_m)=\mathds{1}
	\end{equation} 
	is satisfied.
	
	Upon its exit, the bundle is measured, and the outcome $x_m\in\{1,\dots,d^L\}$ is obtained.
	The measurement on one of the bundles is a projection-valued measure (PVM)~\cite{WM09,Hay2006} $X(x_m)=\ket{x_m}\bra{x_m}$, where
	the outcome $x_m$ is obtained with a probability
	\begin{equation}
		\label{eq:prob_x} P(x_m|\phi,\Phi_{m-1},\rho_{N-m+1}) =\operatorname{tr}\left( X(x_m)K_i(\phi,\Phi_{m-1})\rho_{N-m+1}K_i^\dagger(\phi,\Phi_{m-1})X(x_m)\right),
	\end{equation}
	omitting the tensor product $\prod\limits_{m'\neq m}\otimes\mathds{1}^{(m')}$ on the operators.
	
	We create a shorthand for the $m^\text{th}$ measurement by combining the quantum channel operator and the measurement operator into
	\begin{equation}
		\label{eq:measurement operator} C_{x_m}(\phi,\Phi_{m-1})=X(x_m)K(\phi,\Phi_{m-1}).
	\end{equation}
	The unnormalized state after this step is
	\begin{equation}
		\label{eq:state_update}
		\rho_{N-m}=C_{x_m}(\phi,\Phi_{m-1})\rho_{N-m+1}C_{x_m}^\dagger(\phi,\Phi_{m-1}).
	\end{equation}
	As the measurement procedure starts from the input state $\rho_N$, $\rho_{N-m+1}$ is determined by the previous measurement outcomes, and so we rewrite the notation $P(x_m|\phi,\Phi_{m-1},\rho_{N-m+1})$ as $P(\bm{x}_m|\phi,\Phi_{m-1})$ to indicate the probability of obtaining $x_m$ given the the previous outcomes being $x_1\dots x_{m-1}$.
	
	The most general feedback procedure uses the entire history $\bm{x}_m$ along with the policy $\varrho_N$ to determine the adjustment $\Phi_{m-1}\mapsto\Phi_{m}$. 
	Therefore, the controllable parameter is a function $\Phi_m(\bm{x}_m)$. 
	We initialize $\Phi_0\equiv0$ at the beginning of the procedure, and after each following adjustment the value $\Phi_m$ can be considered as the best guess for the value of $\phi$. The estimate $\wtphi$ is only inferred after all the particles have been measured from the controllable parameter, $\wtphi\equiv\Phi_M\left(\bm{x}_M\right)$, and therefore $\Phi_M$ is an estimator that is an injective function but not necessarily bijective.
	This function takes a string of discrete random variables as input and hence the range of $\Phi_M$ is also discrete.
	The estimate $\wtphi$ is then a discrete approximation of a continuous parameter $\phi\in\left[0, 2\pi\right)$.
	
	\subsection{Adaptive measurement as a decision-making process}\label{subsec:decision}
	An AQEM procedure can be delineated as a decision-making process, where the function $\Phi_{m}(\bm{x}_m)$ makes the decision of what the value of $\Phi_{m+1}$ should be based on the outcomes.    
	If the outcome $x_m$ is the only information used, the feedback is Markovian~\cite{WMW02,WO12}, but we do not restrict our formalism to this case.    
	Considering the adaptive measurement scheme as a decision-making process allows us to visualize the procedure as a decision tree, where each branch represents a single-shot procedure obtaining a unique string $\bm{x}_M$ occurring with probability $P(\bm{x}_M|\phi,\varrho)$.
	This decision tree can also represent a policy, making it possible to calculate an important property of a policy, namely the size, which contributes to the difficulty in devising the optimal adaptive procedure.

    For an AQEM scheme that utilizes $M$ bundles of $L$ $d$-level particles, there are $M$ measurements in a single-shot procedure, each with $d^L$ possible outcomes.
    This process can be represented by a decision tree of depth $M$ with $d^L$ branches stemming from each node (Fig.~\ref{fig:decision tree}).
    Each branch, from root to leaf, corresponds to an adaptive measurement producing a unique string of outcomes $\bm{x}_M$. Given the estimator $\Phi_M$, this leads to an estimate $\wtphi$, which is not necessarily unique.
	
	Assuming the feedback is a deterministic procedure, the probability of obtaining a sequence of $\bm{x}_M$ is determined by the quantum measurement.
	Because each outcome is generated with a probability $P(\bm{x}_m|\phi,\varrho)$, the probability of obtaining $\bm{x}_M$ is then
	\begin{equation}
		\label{eq:prob_outcome}
		P(\bm{x}_M|\phi,\varrho) = \prod_{m=1}^{M}P(\bm{x}_m|\phi,\Phi_{m-1}).
	\end{equation}
	This probability is also the probability for a single-shot adaptive measurement to deliver an estimate $\wtphi=\Phi_M(\bm{x}_M)$.
	
	A decision tree can also represent a policy, providing the rule for mapping $\Phi_{m-1}\mapsto\Phi_m$ according to where on the decision tree the scheme currently resides.
	Hence, the size of the decision tree; i.e., the number of branches, gives the maximum size of the policy. 
	For a measurement procedure using $M$ bundles of $L$ $d$-level particles, the number of branches is 
	\begin{equation}
		\label{eq:policy size}
		\sum\limits_{m=1}^M(d^L)^m = d^L\frac{(d^{N}-1)}{d^L-1}.
	\end{equation}
	That is, the maximum size of a non-Markovian policy scales exponentially with $N$, making a non-Markovian strategy difficult to devise.
	
	\subsection{Imprecision of an adaptive scheme}\label{subsec:imprecision}    
    In this subsection, we discuss the imprecision determined from the decision tree.
    Whereas there are many ways of quantifying the imprecision of a measurement scheme~\cite{DJK15,ZPK12}, 
    all of them can be said to quantify the width of a distribution of the estimate $P(\wtphi|\phi,\varrho)$.
    Using the estimator $\Phi_M(\bm{x}_M)$, we can relate the distribution of the estimate to the distribution of the outcomes $P(\bm{x}_M|\phi,\varrho)$.
    Minimization of the imprecision is then achieved by optimizing the probabilities associated with the branches of the decision tree.
	
\begin{figure}
		\centering
		\includegraphics[scale=0.7]{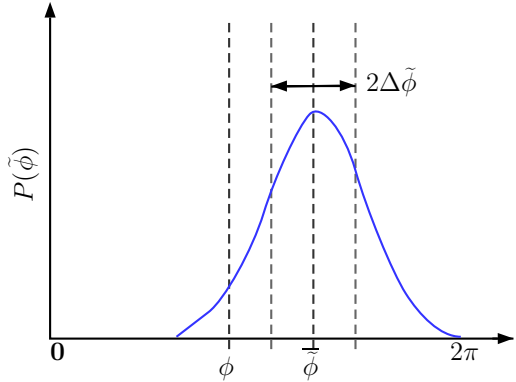}
		\caption{
			Distribution of sampled values of phase estimate~$\wtphi$.
			The actual phase shift~$\phi$, the mean~$\overline\wtphi$
			and the points~$\overline\wtphi\pm\Delta\wtphi$ one standard deviation~$\Delta\wtphi$
			above and below the mean~$\wtphi$ are shown on the abscissa.
		}
		\label{fig:distribution}
	\end{figure}

	The imprecision of a single-shot adaptive measurement can be ascertained from multiple estimations of the same value of $\phi$ (Fig.~\ref{fig:distribution}).
	The data set$\{\wtphi\}$ forms a distribution that peaks at $\bar{\wtphi}$, assuming the distribution is Gaussian. 
	For an unbiased scheme, $\bar{\wtphi}=\phi$, but for small $M$ this is not the case because the estimates have discrete values. 
	The bias of a scheme can be quantified by the difference $\left|\phi-\bar{\wtphi}\right|$.
	
	We define the imprecision to be the spread of the distribution around $\bar{\wtphi}$ regardless of whether the estimate is biased or unbiased.    
	Because of the deterministic relationship between $\bm{x}_M$ and $\wtphi$, we can relate the probability distribution of the outcomes to the probability distribution of the estimate
	\begin{equation}
		\label{eq:PofEstimate_deterministic}
		P\left(\wtphi|\phi,\varrho\right)
		=\sum\limits_{\bm{x}_M\in\Phi_M^{-1}(\wtphi)}P\left(\bm{x}_M|\phi,\varrho\right),
	\end{equation}
	where $\Phi_M^{-1}$ is the inverse function of $\Phi_M$.
	By substituting Eq.~\ref{eq:prob_outcome} in Eq.~\ref{eq:PofEstimate_deterministic}, we obtain a distribution that explicitly depends on the policy. 
	
    From Eq.~\ref{eq:PofEstimate_deterministic}, we have shown that in AQEM minimizing the imprecision is done by optimizing the distribution $P(\bm{x}_M|\phi,\varrho)$. 
    This task is difficult because there are $d^{N}$ possible value of $\bm{x}_M$. 
    Furthermore, each string is obtained through a unique state trajectory, and taking all of the possible trajectories into account in order to compute $P(\wtphi|\phi,\varrho)$ is resource consuming. 
    As a result, generating a feasible policy is a primary challenge for devising an AQEM scheme.
    Before we address this issue, we first discuss the lower bound of an AQEM scheme based on this formalism.
	
	\section{Cram\'{e}r-Rao lower bound for AQEM}\label{sec:CRLB}
	Whether a quantum metrology scheme attains a quantum-enhanced precision depends on its optimal performance, i.e., the minimum imprecision of the scheme.
	The lower bound is defined in parameter estimation by the Cram\'{e}r-Rao lower bound (CRLB), which gives the minimum variance in terms of Fisher information\cite{Oli2014}.
	In this section, we discuss the lower bound for the imprecision of an AQEM scheme using the distribution derived from our decision-making model.
	In particular, we consider the case where the bundles of particles are not entangled, and the lower bound is simplified.
	Because the distribution depends on the policy, the lower bound can only be computed when an optimal policy is given.

	In the classical picture, the lower bound to imprecision $\Delta\wtphi$ arises from imperfection in the measurement device.
	Therefore, in principle, the precision of a classical measurement scheme can be improved indefinitely. 
	When the quantum nature of the measurement scheme is taken into account, the uncertainty principle prevents an improvement to the lower bound beyond that allowed by quantum mechanics.
	To describe the lower bounds for an AQEM scheme, we consider a measurement where there is no imperfection in the measurement scheme and the imprecision is an effect of the quantum-mechanical nature of the input state.
	
	The Cram\'{e}r-Rao lower bound, which is defined as the lower bound for the variance,
		\begin{equation}
		\label{eq:variance}
		\left(\Delta\wtphi\right)^2 = \int\limits_{0}^{2\pi}\text{d}\wtphi P(\wtphi|\phi)\left(\phi-\wtphi\right)^2,
		\end{equation}
	 is derived most generally for multiple-parameter estimation~\cite{Kay93,Zeg15}. Here we consider the case where a single parameter is measured, and the estimate is unbiased.
	Under these conditions, the CRLB becomes
	\begin{equation}    
		\label{eq:DeltaPhiF}
		\Delta\wtphi\geq\frac{1}{\sqrt{F_M(\phi,\varrho)}},
	\end{equation}
	where $F(\phi,\varrho)$ is the Fisher information
	\begin{equation}
		\label{eq:Fisherinformation}
		F_M(\phi,\varrho)
		:=\sum\limits_{\{\bm{x}_M\}}~P(\bm{x}_M|\phi,\varrho)\left(\frac{\partial\log P(\bm{x}_M|\phi,\varrho)}{\partial\phi}\right)^2.
	\end{equation}
	The Fisher information shows the degree for which the outcomes $\{\bm{x}_M\}$ generated from $\phi$ is differentiable from the outcomes generated by $\phi+\text{d}\phi$.
	The information increases as the two sets of outcomes become distinct from one another.
	
	When all the bundles are entangled, the Fisher information cannot be simplified from Eq.~(\ref{eq:Fisherinformation}), although the probability for $\bm{x}_M$ is a product of the probabilities for $x_m$.
	That is because the distribution depends on the history of the outcomes through the back-action on the state.
	If the bundles are independent of one another, the back-action does not affect the subsequent measurements.
	In this case, the Fisher information can be simplified to
	\begin{equation}
		F_M(\phi,\varrho)=\sum\limits_{m=1}^{M}F_1(\phi,\Phi_m).
	\end{equation}
	
	If $F_1(\phi,\Phi_m)$ are identical for all $m\in\left\{1,\dots,M\right\}$, the information become $F_M(\phi,\varrho)=MF_1(\phi,\Phi_1)$, and the SQL is derived if $MF_1(\phi,\Phi_m)\propto N$.
	However, this condition is not satisfied in the presence of a feedback control, which changes the distribution of $P(\bm{x}_m|\phi,\Phi_m)$ in subsequent measurements.
	A statement about the lower bound can only be made when the optimal policy is known.
	To this end, we turn to the approach of reinforcement learning in order to generate a successful policy.
	
	\section{Reinforcement learning for AQEM}\label{sec: learning algorithm}
	In this section, we discuss the use of reinforcement learning for controlling an AQEM scheme.
	This approach has been used to attain better than SQL precision for adaptive phase estimation\cite{HS10,HS11b,LCPS13} with limitation in $N$.
	The algorithm is employed in the training stage, where the unknown parameter $\phi$ is substituted by a set of known parameters, and then a policy generated in this stage is used by the processing unit for the measurements afterward.
	
	Generating a successful policy is an act of optimizing, aiming to obtain a policy $\varrho$ that minimizes the imprecision of an AQEM scheme.
	In general, the imprecision is not necessarily the same for all value of $\phi$, 
	and so the goal of AQEM should be reinstated as minimizing the average imprecision over all possible value of $\phi$.
	As $\phi$ is in a continuous domain, this goal adds difficulty to the optimization process already dealing with a large number of possibilities.
	In the framework of learning algorithm, this issue is handled by sampling a finite set of $\phi$ from the domain~\cite{Mur12}, thereby providing us with a way of performing the optimization without having to take all possible values of $\phi$ into account.
	
	A reinforcement learning algorithm is a type of learning algorithm designed for a learning problem that includes stochastic interactions between the agent (processing unit) and the environment (measurement).
	The goal of the algorithm is to find a policy that maximizes the agent's performance of a task, quantified by an objective function, on a set of given inputs.
	Because of the stochastic interactions, a training set for this type of learning problem does not have an input-output pair structure, as in supervised learning~\cite{Mit97}.
	The type of learning problems that fit this description can be found in classical feedback control~\cite{SBW92} and game playing~\cite{SSM07,SHM+16}.
	
	At the core of a reinforcement learning algorithm is an optimization algorithm, which is the part where a successful policy is generated. 
	In particular, we consider the heuristic optimization algorithms, which act as search algorithms inspired by processes found in nature~\cite{WG04}. 
	The algorithm searches for a feasible policy through trials and errors instead of using the gradient of the objective function to anticipate the direction towards a successful policy~\cite{KRK+05}.
	This approach gives us an advantage of optimizing independently from the AQEM dynamics~\cite{Bis06}.
	In principle, the simulation of an AQEM scheme can be replaced with an experimental setup where the dynamics include noise and loss is not fully described.
	
	With this approach, we can generate a policy that delivers the lower bound for an AQEM scheme.
	In particular, we are interested in observing the lower bound of an AQEM using non-entangled state.
	As the dynamics of the adaptive scheme is complex, we turn to the numerical simulation of a well-studied estimation problem: adaptive phase estimation.
	
	\section{Adaptive phase estimation}\label{sec: phase estimation}
	In this section, we outline the adaptive phase estimation scheme and formulate it for the reinforcement learning algorithm.
	In this scheme, an $N$-photon state is divided into single-photon pulses injected into a Mach-Zehnder interferometer.
	The feedback control is Markovian, which reduces the size of the policy to a linear scaling of $N$, making the problem tractable for a heuristic optimization algorithm.
	The learning algorithm we applied uses a global optimization algorithm as the problem is non-convex and a greedy algorithm has been shown to fail for $N>20$~\cite{PWZS16}.
	
	An adaptive phase estimation scheme has been proposed that has been shown to be able to attain quantum-enhanced precision~\cite{BW00}.
	This scheme uses an $N$-photon entangled state given in the basis of the two input modes of a Mach-Zehnder interferometer as
	\begin{equation}
		\ket{\psi}_N
		=
		\left(\frac{N}{2}+1\right)^{-1/2}
		\sum\limits_{n,k=0}^{N}\sin\left(\frac{k+1}{N+2}\pi\right)\text{e}^{\text{i}\pi (k-n)/2}
		d_{n-N/2,k-N/2}^{N/2}\left(\frac{\pi}{2}\right)
		\ket{n,N-n}.
		\label{eq: sine state}
	\end{equation}
	The variable $n$ in the basis denotes the total number of the photons being injected into the first port of the interferometer regardless of the order, hence making the basis and the state permutationally symmetric.
	We refer to this state as the \textit{sine state}, and it is known to be an optimal state for phase estimation using a finite number of particles~\cite{SP90,LP96,MC09}.    
	
	The sine state gives us a comparison to the case when a non-entangled state is used.
	We choose as an example of a non-entangled state the state
	\begin{equation}
		\label{eq:product state}
		\ket{\psi}_N=\ket{1,0}^{\otimes N},
	\end{equation}
	which is a sequence of single photon injected into one of the input modes.
	This state is one of the many possible $N$-photon product states, another possibility being one of the Bell states.
	In a preliminary test, we observe no improvement in using a Bell state from using this simple product state.    
	
	The photons are injected into a Mach-Zehnder interferometer one at a time. 
	The interferometer contains an unknown parameter $\phi$ in one arm and a controllable parameter $\Phi$ in another.
	After a photon exits the interferometer, it is detected by single-photon detectors, and the output path is recorded as the outcome $x_m\in\{0,1\}$.
	This information is then used by the processing unit to adjust $\Phi$.
	
	In this scheme, the feedback procedure is Markovian, meaning that only $x_m$ is used by the processing unit and not the entire history of outcomes.
	The rule of updating $\Phi$ is given as 
	\begin{equation}
		\Phi_m=\Phi_{m-1}-(-1)^{x_m}\Delta_m.
	\end{equation}
	As a result of imposing this heuristics, the policy, which is represented by a binary decision tree of depth $N$, can be written as a vector,
	\begin{equation}
		\label{eq:linear policy}
		\varrho=\left(\Delta_1,\Delta_2,\dots,\Delta_N\right),
	\end{equation}
	and so the policy scales linearly with $N$ when it otherwise would have scaled as $2^N$.
	This reduction of size reduces the dimension of the space that the optimization algorithm searches,
	hence making the problem computationally tractable.
	
	The scheme is trained using a set of $\phi$ with the size $K=10N^2$, each $\phi$ sampled uniformly randomly from the domain $\left[0,2\pi\right)$.
	The random selection is so that the policy does not overfit the training data~\cite{Die95}. 
	In other words, the policy delivers similar imprecision when applied to the value of $\phi$ not included in the training set.
	The imprecision of the scheme is estimated using Holevo variance~\cite{WK97},    
	\begin{align}
		V_H &= S^{-2} - 1\label{eq:holevo}\\
		S&=\left|\sum\limits_{k=1}^{K}\frac{\exp\left[\text{i}(\phi_k-\Phi_k)\right]}{K}\right|
		\label{eq:sharpness}.
	\end{align}
	We do not use the variance as defined in Eq.~(\ref{eq:variance}) due to $\phi$ being a periodic parameter.
	The distribution $\Phi(\wtphi|\phi)$ has a $2\pi$ periodicity, and without taking this into account the variance is well-defined.
	The size $K$ is chosen such that Eq.~(\ref{eq:holevo}) converges to the true variance, and adding more sample changes the estimated imprecision within an acceptable bound.
	
	We construct a learning algorithm based on differential evolution (DE)~\cite{SP97}, which is an optimization algorithm inspired by biological evolution.
	This algorithm has been observed to perform better than other heuristic optimization algorithms such as particle swarm optimization~\cite{KE95} and other evolutionary algorithms in finding a policy for 100-dimensional space~\cite{VT04,PSL05}.
	In a previous implementation, DE has delivered successful policies up to $N=93$~\cite{LCPS13}, surpassing the use of particle swarm optimization~\cite{HS10,HS11b} which show stagnation in imprecision for $N>45$. 
	We modify the algorithm such that enough computational resource is directed towards finding a feasible policy for $N>93$\footnote{\url{http://panpalitta.github.io/phase_estimation/}}.
	We are able to observe power-law scaling in imprecision up to $N=100$~\cite{PWZS16}, where numerical error and restriction in computation time make the simulation for $N>100$ currently impractical.
	
	\section{Results and discussion}\label{sec:results}    
	In this work, we apply a learning algorithm to adaptive phase estimation, and using the generated policy we show two sets of results.
	We show that the learning algorithm is able to generate policies that deliver quantum-enhanced precision when an entangled state is used.
	The same algorithm is applied to adaptive phase measurement using non-entangled state.
	
	\begin{figure}
		\centering
		\includegraphics[scale=0.5]{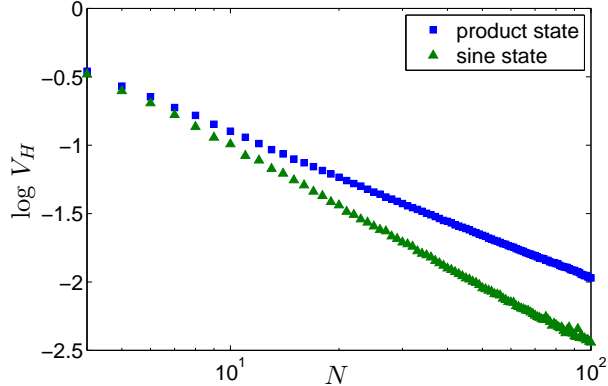}
		\caption{Logarithm of Holevo variances generated using a non-entangled state (blue $\blacksquare$)  and an entangled state (green $\blacktriangle$).}
		\label{fig:result}
	\end{figure}
	We apply the reinforcement learning algorithm to adaptive phase estimation that utilizes $N=\{4,5,\dots,100\}$, and the linear regression of the log-log plot is computed to obtain the power-law scaling (Fig.~\ref{fig:result}). The scaling observed from the non-entangled state is $V_H \propto N^{-1.071}$, which closely adheres to the SQL. A slight improvement can be attributed to the limit of $N=100$, which is far from $N\rightarrow\infty$ assumed for SQL.
	We find that using the sine state leads to a scaling $V_H \propto N^{-1.421}$, which exceeds the SQL. Hence, entanglement between the particles is a crucial part that leads to AQEM being able to attain quantum-enhanced precision. 
	
	\section{Conclusion}\label{sec:conclusion}
	
	In this work, we formalize the description of an AQEM scheme that divides its resource equally over $M$ measurements and model the procedure as a decision-making process.
	Using this framework, we are able to derive the imprecision and the CRLB that take into account the feedback procedure.
	Because of the difficulty in optimizing the policy, we turn to reinforcement learning as an approach for generating a policy that can achieve the quantum-enhanced precision.
	We apply a reinforcement learning algorithm modified to attain power-law scaling up to $N=100$ to an adaptive interferometric phase estimation scheme and is able to show that the quantum-enhanced precision is attained when entanglement is involved. 
	Otherwise, the optimal scaling approaches the SQL. 
	
	
	\section*{ACKNOWLEDGMENTS} 
	This work is financially supported in part by NSERC and AITF. Other sources of funding include ERC (Consolidator Grant QITBOX), MINECO (Severo Ochoa grant SEV-2015-0522 and FOQUS), Generalitat de Catalunya (SGR 875), Fundaci\'o Privada Cellex, and the 1000 Talent Plan.
	The computational work is enabled by the support of WestGrid (www.westgrid.ca) and Compute Canada Calcul Canada (www.computecanada.ca).
	
	\bibliography{aqem} 

\begin{thebibliography}{10}

\bibitem{GLM11}
Giovannetti, V., Lloyd, S., and Maccone, L., ``Advances in quantum metrology,''
  {\em Nat. Photon.}~{\bf 5},  222--229 (Mar 2011).

\bibitem{TA14}
T\'{o}th, G. and Apellaniz, I., ``Quantum metrology from a quantum information
  science perspective,'' {\em J. Phys. A: Math. Theor.}~{\bf 47},  424006 (Oct
  2014).

\bibitem{BS13}
Borregaard, J. and S\o{}rensen, A.~S., ``Near-{H}eisenberg-limited atomic
  clocks in the presence of decoherence,'' {\em Phys. Rev. Lett.}~{\bf 111},
  090801 (Aug 2013).

\bibitem{CSSM14}
Chua, S. S.~Y., Slagmolen, B. J.~J., Shaddock, D.~A., and McClelland, D.~E.,
  ``Quantum squeezed light in gravitational-wave detectors,'' {\em Class.
  Quantum Grav.}~{\bf 31},  183001 (Sep 2014).

\bibitem{JKF+09}
Jones, J.~A., Karlen, S.~D., Fitzsimons, J., Ardavan, A., Benjamin, S.~C.,
  Briggs, G. A.~D., and Morton, J. J.~L., ``Magnetic field sensing beyond the
  standard quantum limit using 10-spin {NOON} states,'' {\em Science}~{\bf
  324},  1166--1168 (May 2009).

\bibitem{ZPK12}
Zwierz, M., {P\'erez-Delgado}, C.~A., and Kok, P., ``Ultimate limits to quantum
  metrology and the meaning of the {H}eisenberg limit,'' {\em Phys. Rev.
  A}~{\bf 85},  042112 (Apr 2012).

\bibitem{Cav81}
Caves, C.~M., ``Quantum-mechanical noise in an interferometer,'' {\em Phys.
  Rev. D}~{\bf 23},  1693--1708 (Apr 1981).

\bibitem{AAA+16}
Abbott, B.~P. et~al., ``Observation of gravitational waves from a binary black
  hole merger,'' {\em Phys. Rev. Lett.}~{\bf 116},  061102 (Feb 2016).

\bibitem{BIWH96}
Bollinger, J.~J., Itano, W.~M., Wineland, D.~J., and Heinzen, D.~J., ``Optimal
  frequency measurements with maximally correlated states,'' {\em Phys. Rev.
  A}~{\bf 54},  R4649--R4652 (Dec 1996).

\bibitem{HBB+07}
Higgins, B.~L., Berry, D.~W., Bartlett, S.~D., Wiseman, H.~M., and Pryde,
  G.~J., ``Entanglement-free {H}eisenberg-limited phase estimation,'' {\em
  Nature}~{\bf 450},  393--396 (Sep 2007).

\bibitem{WBB+09}
Wiseman, H.~M., Berry, D.~W., Bartlett, S.~D., Higgins, B.~L., and Pryde,
  G.~J., ``Adaptive measurements in the optical quantum information
  laboratory,'' {\em IEEE J. Sel. Top. Quantum Electron.}~{\bf 15},  1661--1672
  (Nov 2009).

\bibitem{OIO+12}
Okamoto, R., Iefuji, M., Oyama, S., Yamagata, K., Imai, H., Fujiwara, A., and
  Takeuchi, S., ``Experimental demonstration of adaptive quantum state
  estimation,'' {\em Phys. Rev. Lett.}~{\bf 109},  130404 (Sep 2012).

\bibitem{RPH2015}
Roy, S., Petersen, I.~R., and Huntington, E.~H., ``Robust adaptive quantum
  phase estimation,'' {\em New J. Phys.}~{\bf 17},  063020 (Jun 2015).

\bibitem{Wis95}
Wiseman, H.~M., ``Adaptive phase measurements of optical modes: Going beyond
  the marginal {$Q$} distribution,'' {\em Phys. Rev. Lett.}~{\bf 75},
  4587--4590 (Dec 1995).

\bibitem{LDT14}
{von der Linden}, W., Dose, V., and {von Toussaint}, U.,  [{\em Bayesian
  Parameter Estimation}{\nolinebreak\hspace{0.1em}]}, ch.~14, Cambridge
  University Press, Cambridge, MA, 1~ed. (Jun 2014).

\bibitem{BWB01}
Berry, D.~W., Wiseman, H.~M., and Breslin, J.~K., ``Optimal input states and
  feedback for interferometric phase estimation,'' {\em Phys. Rev. A}~{\bf 63},
   053804 (May 2001).

\bibitem{Oli2014}
Olive, D.~J.,  [{\em Point Estimation {II}}{\nolinebreak\hspace{0.1em}]},
  157--182, Springer, Cham (2014).

\bibitem{HS10}
Hentschel, A. and Sanders, B.~C., ``Machine learning for precise quantum
  measurement,'' {\em Phys. Rev. Lett.}~{\bf 104},  063603 (Feb 2010).

\bibitem{PWZS16}
Palittapongarnpim, P., Wittek, P., Zahedinejad, E., and Sanders, B.~C.,
  ``Learning in quantum control: High-dimensional global optimization for noisy
  quantum dynamics,'' {\em arXiv:1607.03428}  (Jul 2016).

\bibitem{HBZW12}
Hall, M. J.~W., Berry, D.~W., Zwierz, M., and Wiseman, H.~M., ``Universality of
  the {H}eisenberg limit for estimates of random phase shifts,'' {\em Phys.
  Rev. A}~{\bf 85},  041802 (Apr 2012).

\bibitem{Bra92}
Braunstein, S.~L., ``Quantum limits on precision measurements of phase,'' {\em
  Phys. Rev. Lett.}~{\bf 69},  3598--3601 (Dec 1992).

\bibitem{DJK15}
Demkowicz-Dobrza\'{n}ski, R., Jarzyna, M., and Ko\l{}ody\'{n}ski{}, J.,
  ``Quantum limits in optical interferometry,'' in [{\em Progress in
  Optics}{\nolinebreak\hspace{0.1em}]},  Wolf, E., ed.,  {\bf 60}, ch.~4,  345
  -- 435, Elsevier, Amsterdam (2015).

\bibitem{PS09}
Pezz\'e, L. and Smerzi, A., ``Entanglement, nonlinear dynamics, and the
  {H}eisenberg limit,'' {\em Phys. Rev. Lett.}~{\bf 102},  100401 (Mar 2009).

\bibitem{WO12}
{van Otterlo}, M. and Wiering, M.,  [{\em Reinforcement Learning and {M}arkov
  Decision Processes}{\nolinebreak\hspace{0.1em}]}, vol.~12 of {\em Adaptation,
  Learning, and Optimization},  3--42, Springer, Berlin (2012).

\bibitem{Bra94}
Braunstein, S.~L., ``Some limits to precision phase measurement,'' {\em Phys.
  Rev. A}~{\bf 49},  69--75 (Jan 1994).

\bibitem{Hol12}
Holevo, A.~S., ``Quantum evolutions and channels,'' in [{\em Quantum Systems,
  Channels, Information: A Mathematical
  Introduction}{\nolinebreak\hspace{0.1em}]},  {\em De Gruyter Studies in
  Mathematical Physics} {\bf 16}, monograph Quantum evolutions and channels,
  103--131, Walter de Gruyter, Berlin (dec 2012).

\bibitem{YMK86}
Yurke, B., McCall, S.~L., and Klauder, J.~R., ``{SU}(2) and {SU}(1,1)
  interferometers,'' {\em Phys. Rev. A}~{\bf 33},  4033--4054 (Jun 1986).

\bibitem{WM09}
Wiseman, H.~M. and Milburn, G.~J.,  [{\em Quantum Measurement and
  Control}{\nolinebreak\hspace{0.1em}]}, Cambridge University Press, Cambridge,
  MA (2009).

\bibitem{Hay2006}
Hayashi, M.,  [{\em Mathematical Formulation of Quantum
  Systems}{\nolinebreak\hspace{0.1em}]},  9--25, Springer, Berlin (2006).

\bibitem{WMW02}
Wiseman, H.~M., Mancini, S., and Wang, J., ``{B}ayesian feedback versus
  {M}arkovian feedback in a two-level atom,'' {\em Phys. Rev. A}~{\bf 66},
  013807 (Jul 2002).

\bibitem{Kay93}
Kay, S.~M.,  [{\em {C}ramer-{R}ao Lower Bound}{\nolinebreak\hspace{0.1em}]},
  ch.~3, Prentice-Hall, Upper Saddle River, NJ (1993).

\bibitem{Zeg15}
Zegers, P., ``Fisher information properties,'' {\em Entropy}~{\bf 17},  4918
  (Jul 2015).

\bibitem{HS11b}
Hentschel, A. and Sanders, B.~C., ``Efficient algorithm for optimizing adaptive
  quantum metrology processes,'' {\em Phys. Rev. Lett.}~{\bf 107},  233601 (Nov
  2011).

\bibitem{LCPS13}
Lovett, N.~B., Crosnier, C., Perarnau-Llobet, M., and Sanders, B.~C.,
  ``Differential evolution for many-particle adaptive quantum metrology,'' {\em
  Phys. Rev. Lett.}~{\bf 110},  220501 (May 2013).

\bibitem{Mur12}
Murphy, K.~P.,  [{\em Machine Learning: a Probabilistic
  Perspective}{\nolinebreak\hspace{0.1em}]}, MIT Press, Cambridge, MA (2012).

\bibitem{Mit97}
Mitchell, T.~M.,  [{\em Machine Learning}{\nolinebreak\hspace{0.1em}]},
  McGraw-Hill, New York, NY, 1~ed. (1997).

\bibitem{SBW92}
Sutton, R.~S., Barto, A.~G., and Williams, R.~J., ``Reinforcement learning is
  direct adaptive optimal control,'' {\em IEEE Control Systems}~{\bf 12},
  19--22 (Apr 1992).

\bibitem{SSM07}
Silver, D., Sutton, R., and M\"{u}ller, M., ``Reinforcement learning of local
  shape in the game of {Go},'' in [{\em 20th Int. Joint Conf.
  Artif.}{\nolinebreak\hspace{0.1em}]},  {\em IJCAI'07},  1053--1058, Morgan
  Kaufmann, San Francisco, CA (2007).

\bibitem{SHM+16}
Silver, D., Huang, A., Maddison, C.~J., Guez, A., Sifre, L., van~den Driessche,
  G., Schrittwieser, J., Antonoglou, I., Panneershelvam, V., Lanctot, M.,
  Dieleman, S., Grewe, D., Nham, J., Kalchbrenner, N., Sutskever, I.,
  Lillicrap, T., Leach, M., Kavukcuoglu, K., Graepel, T., and Hassabis, D.,
  ``Mastering the game of {Go} with deep neural networks and tree search,''
  {\em Nature}~{\bf 529},  484--489 (Jan 2016).

\bibitem{WG04}
Winker, P. and Gilli, M., ``Applications of optimization heuristics to
  estimation and modelling problems,'' {\em Comput. Stat. Data Anal.}~{\bf
  47}(2),  211 -- 223 (2004).

\bibitem{KRK+05}
Khaneja, N., Reiss, T., Kehlet, C., Schulte-Herbr\"{u}ggen, T., and Glaser,
  S.~J., ``Optimal control of coupled spin dynamics: design of {NMR} pulse
  sequences by gradient ascent algorithms,'' {\em J. Magn. Reson.}~{\bf
  172}(2),  296--305 (2005).

\bibitem{Bis06}
Bishop, C.~M.,  [{\em Pattern Recognition and Machine
  Learning}{\nolinebreak\hspace{0.1em}]}, Springer, Singapore (2006).

\bibitem{BW00}
Berry, D.~W. and Wiseman, H.~M., ``Optimal states and almost optimal adaptive
  measurements for quantum interferometry,'' {\em Phys. Rev. Lett.}~{\bf 85},
  5098--5101 (Dec 2000).

\bibitem{SP90}
Summy, G.~S. and Pegg, D.~T., ``Phase optimized quantum states of light,'' {\em
  Opt. Commun.}~{\bf 77},  75 -- 79 (Jun 1990).

\bibitem{LP96}
Luis, A. and Pe\ifmmode~\check{r}\else \v{r}\fi{}ina, J., ``Optimum phase-shift
  estimation and the quantum description of the phase difference,'' {\em Phys.
  Rev. A}~{\bf 54},  4564--4570 (Nov 1996).

\bibitem{MC09}
Maccone, L. and De~Cillis, G., ``Robust strategies for lossy quantum
  interferometry,'' {\em Phys. Rev. A}~{\bf 79},  023812 (Feb 2009).

\bibitem{Die95}
Dietterich, T., ``Overfitting and undercomputing in machine learning,'' {\em
  ACM Comput. Surv.}~{\bf 27},  326--327 (Sept. 1995).

\bibitem{WK97}
Wiseman, H.~M. and Killip, R., ``Adaptive single-shot phase measurements: A
  semiclassical approach,'' {\em Phys. Rev. A}~{\bf 56},  944--957 (Jul 1997).

\bibitem{SP97}
Storn, R. and Price, K., ``Differential evolution: A simple and efficient
  heuristic for global optimization over continuous spaces,'' {\em J. Global
  Optim.}~{\bf 11},  341--359 (Dec 1997).

\bibitem{KE95}
Kennedy, J. and Eberhart, R., ``Particle swarm optimization,'' in [{\em IEEE
  Int. Conf. Neural Networks}{\nolinebreak\hspace{0.1em}]},   {\bf 4},
  1942--1948, IEEE, Piscataway, NJ (Nov 1995).

\bibitem{VT04}
Vesterstrom, J. and Thomsen, R., ``A comparative study of differential
  evolution, particle swarm optimization, and evolutionary algorithms on
  numerical benchmark problems,'' in [{\em IEEE C. Evol.
  Computat.}{\nolinebreak\hspace{0.1em}]},   {\bf 2},  1980 -- 1987, IEEE (Jun
  2004).

\bibitem{PSL05}
Price, K.~V., Storn, R.~M., and Lampinen, J.~A., ``Benchmarking differential
  evolution,'' in [{\em Differential Evolution}{\nolinebreak\hspace{0.1em}]},
  {\em Natural Computing Series},  135--187, Springer, Berlin (2005).

\end{thebibliography}
	\bibliographystyle{spiebib} 
\end{document}